\documentclass[submission, Proceedings]{SciPost}
\usepackage{amssymb}

\newcommand{\be}{\begin{equation}}
\newcommand{\ee}{\end{equation}}
\newcommand{\bea}{\begin{eqnarray}}
\newcommand{\eea}{\end{eqnarray}}
\newcommand{\lbl}{\label}

\newcommand{\nn}{\nonumber}

\newcommand{\borel}{\mathcal{B}^{[\rho]}(\sigma)}

\definecolor{darkgreen}{rgb}{0.0, 0.4, 0.0}

\begin{document}

\begin{center}{\Large \textbf{ {\begin{boldmath} Determining $\alpha_s$ from hadronic $\tau$  decay:\\ the {pitfalls}  of {truncating the OPE}\end{boldmath}}}
}\end{center}

\begin{center}
D. Boito\textsuperscript{1},
M. Golterman\textsuperscript{2},
K. Maltman\textsuperscript{3,4},
S. Peris\textsuperscript{5,*}
\end{center}

\begin{center}
{\bf 1} Instituto de F\'{\i}sica de S\~{a}o Carlos, {Universidade} de S\~{a}o Paulo, Brazil
\\
{\bf 2} Department of Physics and Astronomy, San Francisco State University, USA
\\
{\bf 3} Department of Mathematics and Statistics, York University, Canada
\\
{\bf 4} CSSM,University of Adelaide, Adelaide, Australia
\\
{\bf 5} Department of Physics and IFAE-BIST, Universitat Aut\`{o}noma de Barcelona, Spain
\\
{*peris@ifae.es}
\end{center}

\begin{center}
\today
\end{center}

\definecolor{palegray}{gray}{0.95}
\begin{center}
\colorbox{palegray}{
  \begin{tabular}{rr}
  \begin{minipage}{0.05\textwidth}
    \includegraphics[width=8mm]{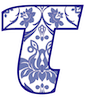}
  \end{minipage}
  &
  \begin{minipage}{0.82\textwidth}
    \begin{center}
    {\it Proceedings for the 15th International Workshop on Tau Lepton Physics,}\\
    {\it Amsterdam, The Netherlands, 24-28 September 2018} \\
    \href{https://scipost.org/SciPostPhysProc.1}{\small \sf scipost.org/SciPostPhysProc.Tau2018}\\
    \end{center}
  \end{minipage}
\end{tabular}
}
\end{center}


\section*{Abstract}
{\bf
\begin{boldmath}
We discuss sum-rule determinations of $\alpha_s$
{from} non-strange hadronic {$\tau$-decay} data. We investigate, in particular, the reliability of {the}
assumptions {underlying the ``truncated OPE strategy,'' which specifies a certain
treatment of non-perturbative contributions, and which was employed in Refs.~\cite{Pich,Aleph, Opal}.  Here, we test this strategy by applying the strategy to the $R$-ratio obtained from
$e^+e^-$  data, which extend beyond the $\tau$ mass, and{, based on the outcome of these tests, we} demonstrate the failure of this
strategy.}
{
{We then present a brief overview of} new results on the form
of {duality-violating non-perturbative} contributions, which are conspicuously present in the experimentally determined spectral functions.
As we show, with the current precision claimed for the extraction of
$\alpha_s$, including a representation of {duality violations} is unavoidable if one
wishes to avoid  uncontrolled theoretical errors.}
\end{boldmath}}

\vspace{10pt}
\noindent\rule{\textwidth}{1pt}
\tableofcontents\thispagestyle{fancy}
\noindent\rule{\textwidth}{1pt}
\vspace{10pt}
\section{Introduction}
\label{sec:intro}
As is well known, the determination of $\alpha_s$ from finite-energy
sum-rule (FESR) analyses of hadronic {$\tau$-decay} data provides
one of the most precise determinations of
$\alpha_s$. Because of its low scale, this determination, moreover,
plays an important role in testing the evolution of the strong
coupling predicted by QCD. In this paper we pull back the curtain
on, and subject to further scrutiny, certain issues {and} subtleties
connected with the treatment of non-perturbative effects;
issues which the precision now claimed for these determinations
makes it important to understand in more quantitative detail.

In what follows, we first demonstrate that certain highly non-trivial
assumptions made in treating non-perturbative contributions in
common implementations of the {FESR} analysis framework can be tested (and shown to fail) via analogous analyses of electromagnetic
(EM) hadroproduction cross-sections, which, unlike hadronic {$\tau$-decay} distributions, are not kinematically restricted to
hadronic invariant-squared-masses $s\leq m_\tau^2$.
These observations {imply} that {$\tau$-decay} analyses
cannot avoid employing weighted spectral integrals with variable upper
endpoints, $s_0\leq m_\tau^2$. Given that significant
duality violations (DVs) are clearly observed in the experimental
differential non-strange hadronic {$\tau$-decay} distributions, this
necessitates providing estimates for the size of residual DV effects,
which in turn necessitates the use of models for the DV components of the
hadronic spectral functions. Recent progress in determining the
form of these components expected in QCD, relevant to carrying out such
analyses, is then also reviewed.

In the Standard Model, defining
\begin{equation}
R_{ud;V/A} \equiv {\frac{\Gamma [\tau \rightarrow \nu_\tau
\, {\rm hadrons}_{ud;V/A}\, (\gamma )]}{ \Gamma [\tau^- \rightarrow
\nu_\tau e^- {\bar \nu}_e (\gamma )] }}\ ,
\label{Rtaudef}\end{equation}
one has~\cite{tsai}
\begin{eqnarray}
&&{\frac{dR_{ud;V/A}}{ds}}\, =\, {\frac{12 \pi^2 \, \vert V_{ud}\vert^2
S_{EW}}{m_\tau^2}}\,\left[ w_\tau \left(y_\tau \right)
\, \rho_{ud;V/A}^{(0+1)}(s)
-\, w_L\left(y_\tau\right)\, \rho_{ud;V/A}^{(0)}(s) \right]\ ,
\label{tsaidRds}\end{eqnarray}
where $y_\tau = s/m_\tau^2$, $w_\tau (y)=(1-y)^2(1+2y)$, $w_L(y)=2y(1-y)^2$,
$V_{ud}$ is the $ud$ element of the CKM matrix, $S_{EW}$ is a known
short-distance electroweak correction~\cite{erlersew}, and
$\rho^{(J)}_{ud;V/A}(s)$ are the spectral functions of the $J=0,1$
hadronic vacuum polarizations (HVPs), $\Pi_{ud;V/A}^{(J)}$, of the flavor
$ud$, vector ($V$) and axial-vector ($A$) current-current two-point
functions. The continuum parts of $\rho_{ud;V/A}^{(0)}(s)$
are suppressed by factors of $(m_d\mp m_u)^2$, and hence numerically
negligible, leaving the well-determined {pion-pole} contribution to
$\rho_{ud;A}^{(0)}$ as the only numerically relevant $J=0$ contribution.
The $J=0+1$ sums $\rho_{ud;V/A}^{(0+1)}(s)$ are thus directly determinable
from the experimental $dR_{ud;V/A}/ds$ distributions.

\begin{figure}[t]
\label{fig:cauchy}
\vspace*{4ex}
\begin{center}
\includegraphics*[width=6cm]{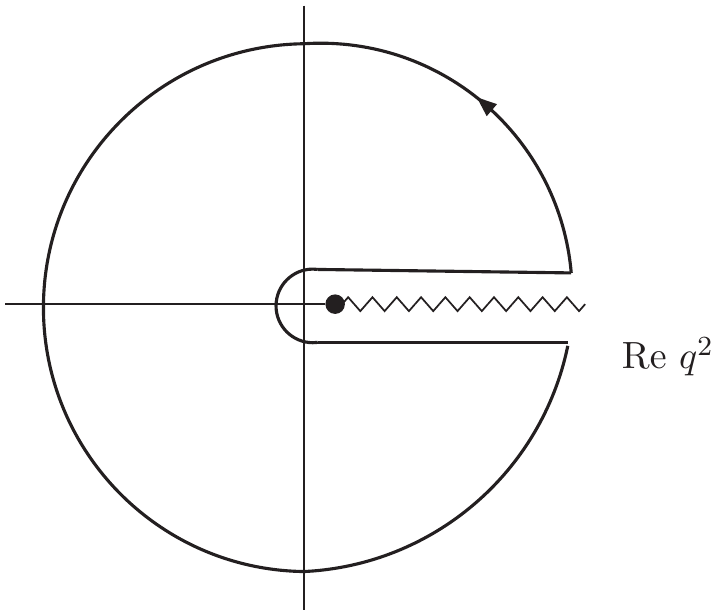}
\end{center}
\begin{quotation}
\caption{\it Contour used in the derivation of
Eq. (\ref{basicfesr}). The cut shown on the positive real
$s=q^2=\, -Q^2$ axis starts at $s=4 m_\pi^2$ for $\Pi_{ud;V}^{(0+1)}$
and $s=9m_\pi^2$ for $\Pi_{ud;A}^{(0+1)}$. $\Pi_{ud;A}^{(0+1)}$,
of course, also has a pole at $s=m_\pi^2$. }
\end{quotation}
\vspace*{-4ex}
\end{figure}

The spectral function combinations $\rho_{ud;V/A}^{(0+1)}(s)$ and
$s\, \rho_{ud;V/A}^{(0)}(s)$ correspond to {HVP} combinations,
$\Pi_{ud;V/A}^{(0+1)}(s)$ and $s \Pi_{ud;V/A}^{(0)}(s)$,
which are free of kinematic singularities. For
any $s_0\leq m_\tau^2$, and any weight $w$ analytic inside and on
$\vert s\vert =s_0$, Cauchy's theorem, applied to the contour
in Fig.~\ref{fig:cauchy}
then ensures the validity of the FESR relation \cite{FESRs}
\begin{eqnarray}
&&\int_0^{s_0}{\frac{ds}{s_0}} \, w(s/s_0)\, \rho_{ud;V/A}^{(0+1)}(s)\, =
{\frac{-1}{2\pi i}}\, \oint_{\vert s\vert =s_0}{\frac{ds}{s_0}}
\, w(s/s_0)\, \Pi_{ud;V/A}^{(0+1)}(s)\ .
\label{basicfesr}\end{eqnarray}

The basic idea of the $\tau${-based} determination of $\alpha_s$ is to
employ experimental {results for} $dR_{ud;V/A}/ds$  on the LHS of
Eq.~\ref{basicfesr} and, for sufficiently large $s_0$, the { Operator Product Expansion} (OPE)
representation of $\Pi_{ud;V/A}^{(0+1)}(s)$ on the RHS. The OPE,
of course, represents only an approximation to $\Pi_{ud;V/A}^{(0+1)}$.
In general, in addition to perturbative (dimension $D=0$ OPE) and
higher dimension non-perturbative OPE condensate contributions,
\begin{equation}
\left[ \Pi_{ud;V/A}(s)\right]_{OPE}^{NP}
=\sum_{D=4,6,8,\cdots}{\frac{C^{V/A}_D}{Q^D}}{\, ,}
\label{npopeform}\end{equation}
with $Q^2=\, -s$, and the $C^{V/A}_D$ effective condensates of
dimension $D$, non-OPE, DV contributions,
$\left[ \Pi_{ud;V/A}^{(0+1)}(s)\right]_{DV}$, defined by
\begin{equation}
\Pi_{ud;V/A}^{(0+1)}(s) \equiv \left[\Pi_{ud;V/A}^{(0+1)}(s)\right]_{OPE}
+\Pi_{ud;V/A}^{DV}(s)\ ,
\label{pidvdefn}\end{equation}
are needed to provide a full representation of $\Pi_{ud;V/A}^{(0+1)}(s)$.

If $m_\tau^2$ were {sufficiently large} that, relative to perturbative
contributions, all non-perturbative contributions (both DV and OPE)
were negligible on the circle $\vert s\vert = m_\tau^2$, the inclusive experimental
non-strange hadronic $\tau$ decay width would provide an immediate
determination of $\alpha_s$. Unfortunately, this is not the case, at
the level of precision desired (and claimed) in current $\tau$-based analyses.

Two key qualitative points should be emphasized regarding non-perturbative
contributions to the RHS of Eq.~(\ref{basicfesr}). First, since the
cut in $\Pi_{ud;V/A}^{(0+1)}$ extends to $s=\infty$ ($z=1/Q^2=0$),
the OPE (an expansion in $z$ about $z=0$) cannot be convergent.
Second, DV contributions to $\Pi_{ud;V/A}^{(0+1)}(s)$,
which are exponentially suppressed for large spacelike $Q^2=\, -s$, are
expected to develop an additional oscillatory {behavior}
{on the Minkowski axis \cite{ShifmanDVs, Regge}. {Such oscillations
are clearly seen in $\rho_{ud;V/A}^{(0+1)(s)}={\frac{1}{\pi}}\,
{\rm Im}\, \Pi_{ud;V/A}^{(0+1)}(s)$} but their properties are not captured
by the OPE. They reflect the incipient presence of resonances as the energy is lowered from the parton-model regime.}

The fact that the OPE is not convergent means that it is not { true}
 that higher dimension OPE contributions to the RHS of
Eq.~(\ref{basicfesr}) scale simply as $\Lambda_{QCD}^D/s_0^{D/2}$ and hence
form a rapidly converging series in $D$ for $s_0\simeq m_\tau^2$.
Assuming, on such
``dimensional'' grounds, {that integrated higher-$D$ OPE contributions {in principle present for a given weight $w$} can be neglected requires experimental justification if one
wishes to avoid incurring unquantifiable systematic
errors.}

\begin{figure}
\label{Pich-RS9}
\includegraphics*[width=7cm]{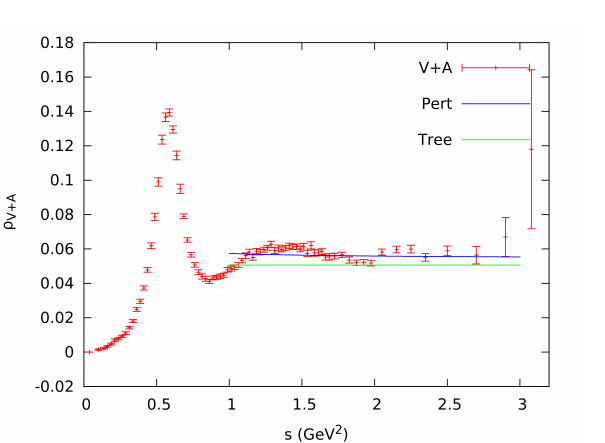}
\hspace{.5cm}
\includegraphics*[width=7cm]{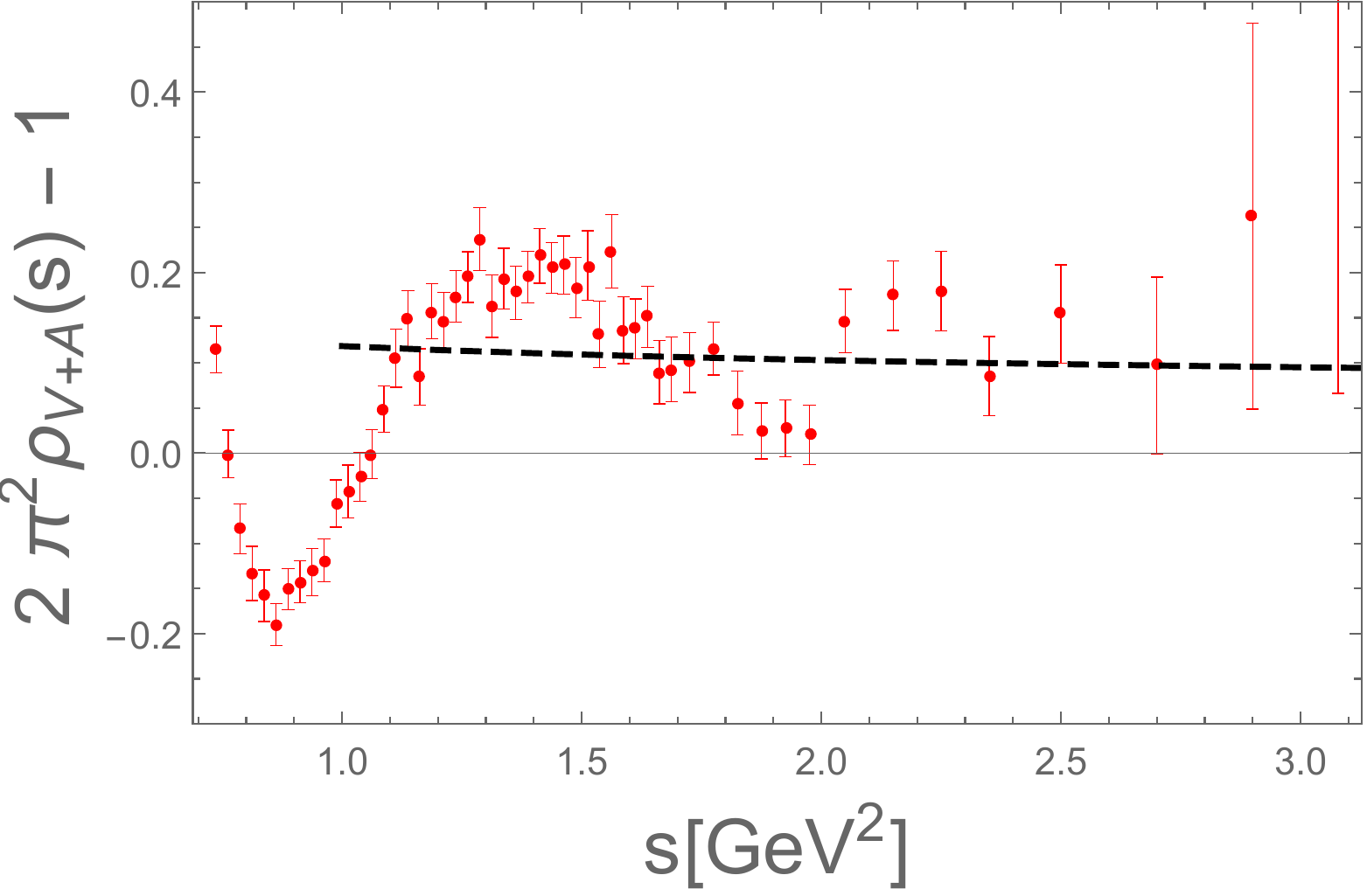}
\begin{quotation}
\caption{{\it Left panel: the $ud$ $V+A$ spectral function,
as shown in Ref. \cite{Pich}. Lower panel: the same (in the normalization of
\cite{appraisal}) but now with the $\alpha_s$-independent parton-model
contribution subtracted.}}
\end{quotation}
\end{figure}

The fact that DV effects are not, in general, negligible in
hadronic $\tau$ decays is evidenced by the size of the
observed DV oscillations in the $V$, $A$ and $V+A$ spectral functions. It
is often argued \cite{Pich}  that DV oscillations are ``small'' for the $V+A$
combination on the basis of plots showing the size of such oscillations
on the scale of the full $V+A$ spectral function, $\rho_{ud;V+A}^{(0+1)}(s)$,
as in the left panel of Fig.~\ref{Pich-RS9}. Such a plot is, however,
highly misleading, since $\rho_{ud;V+A}^{(0+1)}(s)$ contains a large
parton-model contribution completely independent of $\alpha_s$, {{\it i.e.},} of all QCD dynamics. The FESR
determination of $\alpha_s$ is driven entirely by the dynamical,
$\alpha_s$-dependent part of the perturbative contribution to the
weighted spectral integrals, and the relevant measure of the relative
size of perturbative and DV contributions to the spectral functions
entering those integrals, from the point of view of a determination
of $\alpha_s$, is the size of the DV oscillations relative to the
{\it $\alpha_s$-dependent part} of the perturbative representation of
$\rho_{ud;V+A}^{(0+1)}(s)$. The right panel of Fig.~\ref{Pich-RS9}
shows this more relevant comparison. One immediately sees, for
example, that the non-parton-model part of $\rho_{ud;V+A}^{(0+1)}(s)$
is $\simeq 0$ for $s\simeq 2\ {\rm GeV}^2$, indicating that DV and
$\alpha_s$-dependent perturbative contributions are, in fact, equal
in magnitude in this region{{, essentially}  cancelling each other out}. This is also true in the vicinity
of the next DV peak, {where, however, the two contributions combine
constructively, as expected given the oscillatory nature of DVs}. While it {\it is} true that DV oscillations
are smaller for the $V+A$ combination than for the individual $V$ and $A$
spectral functions, this {rather obviously}
does not mean that the $V+A$ oscillations are
small in an absolute sense.

DV contributions, though important in hadronic spectral functions,
certainly for {the range of $s$} accessible in $\tau$ {decays}, may
be suppressed relative to perturbative contributions when one considers
the integrated quantities appearing on the RHSs of Eq.~(\ref{basicfesr}).
From the arguments of Ref.~\cite{pqw}, DV contributions on
$\vert s\vert =s_0$ at intermediate $s_0$ are expected to be localized
to the vicinity of the timelike axis. Given the asymptotic nature of the OPE, and the oscillatory behavior of the DVs, the parametrization \cite{Cata1,Cata2,Oscar}
\begin{equation}
\label{parametrization}
\frac{1}{\pi}\mathrm{Im}\Pi^{DV}_{ud;V/A}(s)=
e^{-\delta_{V/A}-\gamma_{V/A} s} \sin\left(\alpha_{V/A}+
\beta_{V/A} s\right)\ ,
\end{equation}
for $s$ large enough, {represents a very natural} choice.\footnote{Recall how renormalons give rise to a
$e^{-b/\alpha_s}$ behavior from the asymptotic nature of
perturbation theory. Here the expansion parameter
is $1/s$ rather than $\alpha_s$ and $\alpha_s$ is thus {parametrically}
replaced by $1/s$.} In fact, this expression has recently been
confirmed \cite{Regge} under the mild assumption of an asymptotic
Regge behavior for the meson spectrum. We discuss this in more
detail in section \ref{sec:Regge} below.

 The contribution from $\Pi^{DV}_{ud;V/A}(s)$ in Eq.~(\ref{pidvdefn}) to the FESR~(\ref{basicfesr})  can be shown to take the form \cite{Oscar,case}
\begin{equation}
\label{new}
{\frac{-1}{2\pi i}}\, \oint_{\vert s\vert =s_0}{\frac{ds}{s_0}}
\, w(s/s_0)\, \Pi_{ud;V/A}^{DV}(s) = - \int_{s_0}^\infty {\frac{ds}{s_0}}\,w(s/s_0)\,\frac{1}{\pi}\,\mbox{Im}\,
\Pi^{DV}_{ud;V/A}(s)\  ,
\end{equation}
and, using the parametrization (\ref{parametrization}), this form of the DV contributions on the RHS of (\ref{new}) will be very useful in the following discussion.

{ Because of the exponential suppression at large $s$ in Eq.~(\ref{parametrization}),
{and localization of DV contributions to the
vicinity of the timelike axis at intermediate and large $s$,} the use} of ``pinched weights''
$w$ (those with a zero at $s=s_0$) in Eq.~(\ref{basicfesr}) is thus
expected to yield RHSs in the FESRs {(\ref{basicfesr})} in which residual integrated DVs play a reduced
role relative to integrated OPE contributions, with the level of suppression
typically increasing with the degree of pinching (the order of the zero at
$s=s_0$). This general expectation is confirmed empirically~\cite{kmdvs98},
and can also be seen on average, over a range of $s_0$, when one
integrates explicitly the asymptotic DV form (\ref{parametrization}) \cite{case}. While increased pinching increasingly suppresses DVs on
average, given the size of DV contributions to the spectral functions,
and the precision claimed for current versions of the determination
of $\alpha_s$, it remains an open question how large this suppression
is for the doubly and triply pinched weights employed in typical
determinations of $\alpha_s$.

The analyses of Refs.~\cite{Aleph,Opal,Pich},
all implicitly assume the scale $s_0=m_\tau^2$ is high enough that
integrated DVs can be neglected for the doubly and triply pinched
weights entering those analyses. Ref.~\cite{my08} also assumes
integrated DVs can be neglected for the doubly pinched weights
it employs, testing this assumption for self-consistency by
studying the $s_0$ dependence of the resulting fits. In contrast,
the analysis of Ref.~\cite{tau,tau2} employs the model for DV contributions
to the $V$ and $A$ spectral functions of Eq.~(\ref{parametrization}),
and finds a systematic downward shift in the $\alpha_s$ obtained
when this representation of DV effects is included.

Of particular
relevance to the results of Refs.~\cite{my08,Pich}, where
attempts are made to investigate the self-consistency of the assumed
neglect of DVs, are the results of Ref.~\cite{appraisal}, where
the tests employed in Ref.~\cite{Pich} are applied to a model { based on {mock} data}
which accurately matches the experimental $ud$ $V+A$ spectral function
and which has, by construction, a lower input value of $\alpha_s$
{as well as} numerically relevant DV contributions at higher $s$.
It is found that what were hoped to be self-consistency tests in
Ref.~\cite{Pich}, applied to this model, are unable to identify
the presence of the model DV contributions and the lower input
$\alpha_s$ value, establishing that these tests are{, in general,}
insufficient to establish the absence of numerically significant DVs.
A comparison of the results of Refs.~\cite{my08} and \cite{tau}
suggests this same caveat is relevant to assessing the results
of the $s_0$-dependence tests employed in Ref.~\cite{my08}.

{The failure of nominal self-consistency tests of Ref.~\cite{Pich} when applied to the model
described above still leaves open the logical possibility that DV contributions to {the actual
spectral functions} in the region $s>m_\tau^2$
might be smaller than those in the model, leading to
smaller integrated DV contributions in the real world than in the model.   This possibility
can be investigated, at least for the $I=1$ $V$ component, using $e^+e^-$ hadroproduction
cross-section data \cite{knt2018}. The reason is that the $I=1$ component of the EM current is
related by CVC to the charged $I=1$ $V$ current acting in $\tau$ decays.  The predictions of the
$\tau$-based model of Ref.~\cite{appraisal} for the $I=1$ spectral function
in the region $s_0\ge m_\tau^2$, where it
cannot be measured in $\tau$ decays, can then be tested against the $I=1$ component of the
EM spectral function obtained from a $G$-parity based isovector/isoscalar separation of the
$I=0$ and $I=1$ contributions to the EM spectral function. This separation was carried
out in the region {up to $s=4$~GeV$^2$} in Ref.~\cite{bgkmnpt2018}.
The result of the comparison
to the prediction shown in Fig.~5 of Ref.~\cite{bgkmnpt2018} shows good agreement:
the DV oscillations predicted by the $\tau$-decay based model are, indeed, seen in the
$e^+e^-$ data.}

{One} can also use the electroproduction $R(s)$ data to test the ``truncated OPE strategy" which is the foundation for the analysis of Refs. \cite{Pich, Aleph}. The problem with the truncation of the  OPE arises as follows. The spectral
integral for the $s_0=m_\tau^2$ version of the FESR involving the
kinematic weight $w_\tau (y)$, with $y\equiv s/s_0$,
{can directly be determined} from the
inclusive branching fraction for hadronic $\tau$ decays.
This result, however, is insufficient to allow one to determine
$\alpha_s$ since $w_\tau$ has degree $3$, and the theorem of residues then {implies} that the right-hand (theory) side of the $w_\tau$ FESR involves OPE contributions
up to $D=8$. While the $D=4$ contribution is strongly suppressed
by the absence of a term linear in $y$ in $w_\tau(y)$, three OPE
parameters, $\alpha_s$, and the $D=6$ and $8$ effective condensates,
$C_6$ and $C_8$, are still required to fix the theory side. Since
$C_6$ and $C_8$ are not known from external sources, the inclusive
non-strange branching fraction itself cannot provide a determination
of $\alpha_s$.

{A strategy employed to try to get
around this problem \cite{Pich, Aleph,Opal}} is to consider additional $s_0=m_\tau^2$ FESRs
involving new, {higher-degree} weights, with  at least  {the level of pinching of} $w_\tau$. The goal is to use the additional weighted spectral integrals as inputs to an extended multi-weight
analysis in which non-perturbative condensates like $C_6$ and $C_8$
are also fit.

{ This strategy, however, has a {fundamental} shortcoming. If one considers, for
example, using one additional FESR involving a polynomial weight of degree $4$,
that FESR now receives a contribution from a new effective OPE
{$D=10$} condensate, $C_{10}$. Adding a weight with degree
$5$ similarly brings into play a contribution proportional to another
new {$D=12$} condensate,  $C_{12}$, {\it etc}. As long as one
aims to suppress as much as possible residual
integrated DV effects by considering spectral integrals with $s_0=m_\tau^2$
{with at least doubly pinched weights}
only, one has, at every stage, more OPE parameters to fit than weighted
spectral integrals to use in fitting them.

For this strategy to work in practice,
one thus needs to make the strong additional assumption that,
for a set of weights whose maximum degree is $N$, and which, therefore, requires knowledge of OPE condensates up to dimension $D=2N+2$, the OPE can be truncated
{at a dimension smaller than $2N+2$ sufficiently low to leave the number of OPE fit
parameters less than the number of spectral integrals to be used in fitting them. Though this truncation leads to a proper fit {in the statistical sense}, it is really only justified if the asymptotic OPE
series behaves, at the scales of the analysis, as if it were convergent. This approach to
determining
$\alpha_s$ from hadronic $\tau$-decay data,
which we refer to as the ``truncated OPE
strategy,'' has been employed, for example, in Refs.~\cite{Pich,Aleph,Opal}.
}

The truncated OPE strategy is therefore predicated on the assumptions
that $s_0=m_\tau^2$ is large enough that (i) integrated DV contributions
can be neglected for FESRs involving doubly and triply pinched weights
and (ii) the OPE, though asymptotic {at best}, behaves as if it were
rapidly convergent for dimensions up to $2N+2$, where $N$ is the
degree of the highest-degree weight entering the analysis in question.

It is important to stress that integrated DV contributions are
expected to be exponentially damped with increasing $s_0$, and
that integrated {higher-dimension} $D=2k$ OPE contributions scale
as $1/s_0^k$ and hence also decrease, relative to the leading
$D=0$ perturbative contributions, with increasing $s_0$.\footnote{{ Although
the asymptotic nature of the OPE leads to the expectation of a rapid
increase of the condensate contribution with its dimension $D$.} }
{It therefore follows} that, if the assumptions of the truncated OPE strategy
were {valid} for $s_0=m_\tau^2$, they would be even more so for higher $s_0$.
The kinematic {restriction $s_0\leq m_\tau^2$ unfortunately} prevents
this prediction from being tested using {$\tau$-decay} data {but,
fortunately, the {$R$-ratio data obtained from $e^+e^-\to\mbox{hadrons}$
allow for such tests.}}

Analogous EM FESRs,
employing results for $R(s)$ obtained in Ref.~\cite{knt2018},
thus allow us to investigate the reliability of the assumptions
underlying the truncated OPE strategy by applying the same fits to the
correspondingly weighted $s_0=m_\tau^2$ versions of the EM
spectral integrals, and then testing whether the resulting OPE
fit results provide a good representation of the actual EM
spectral integrals for $s_0>m_\tau^2$. This
{investigation} will be the subject of the next section.}

\begin{figure}
\label{optimalwttests}
\vspace*{4ex}
\includegraphics*[width=9cm]{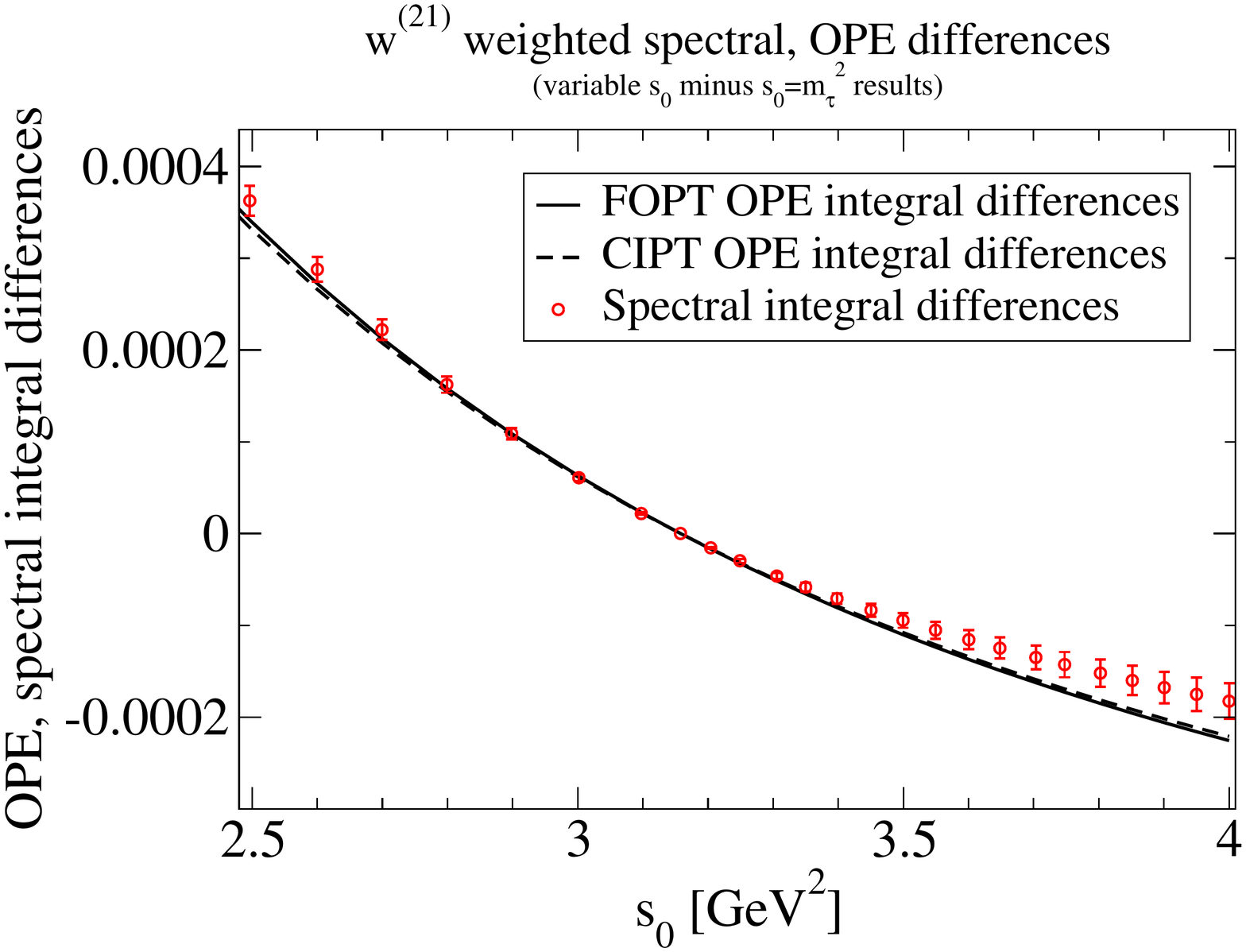}
\hspace{0cm}
\includegraphics*[width=9cm]{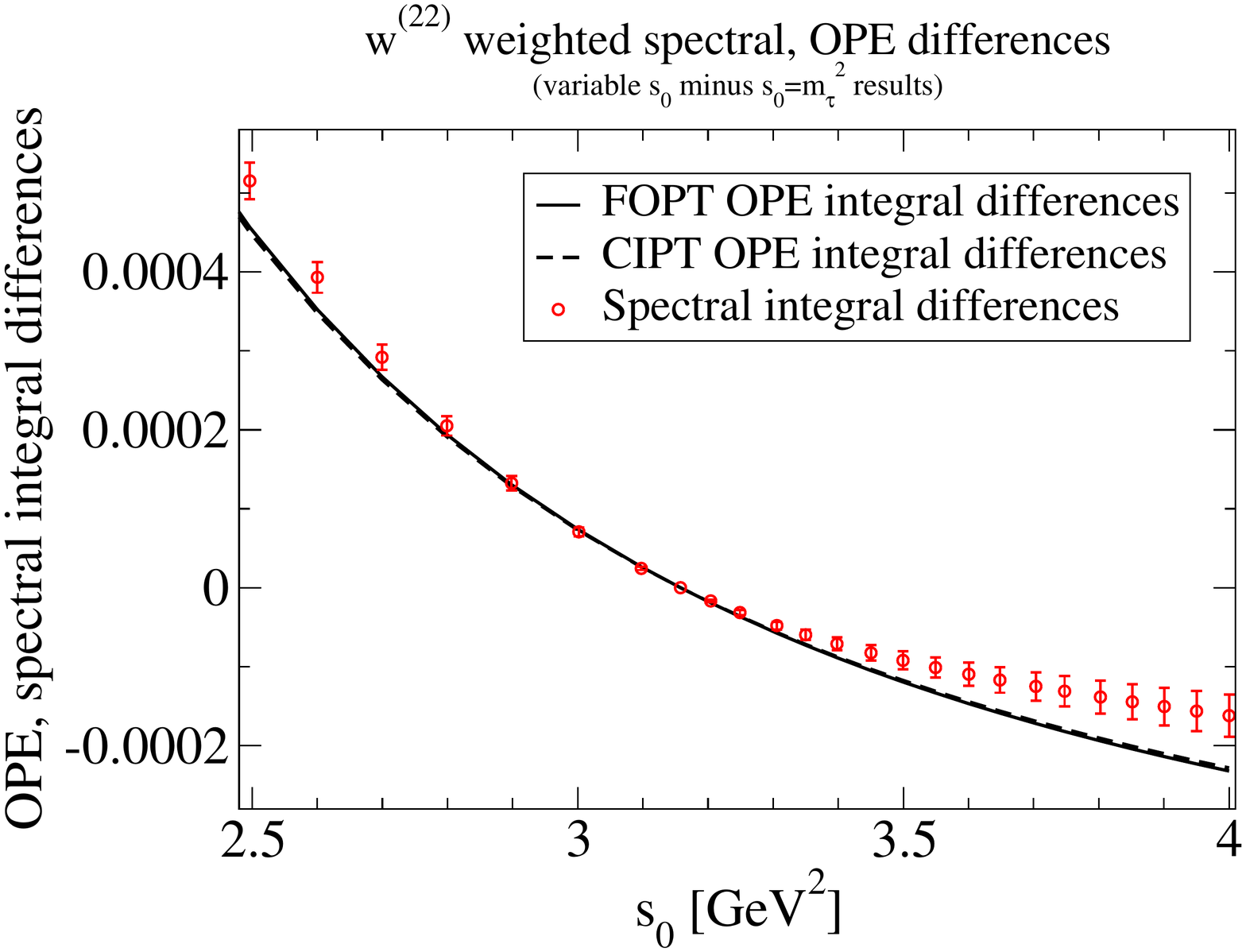}
\vspace{4ex}
\includegraphics*[width=9cm]{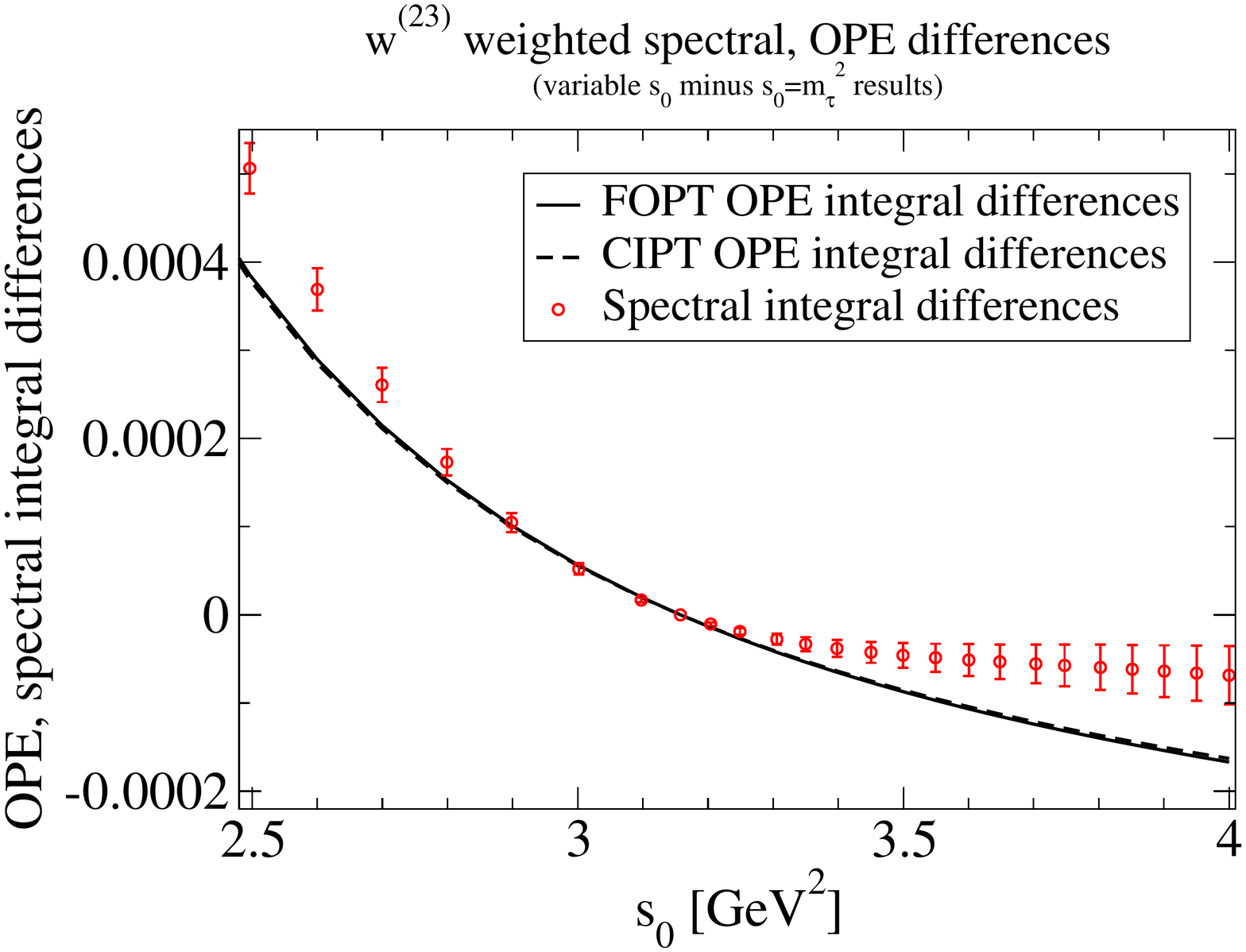}
\hspace{0cm}
\includegraphics*[width=9cm]{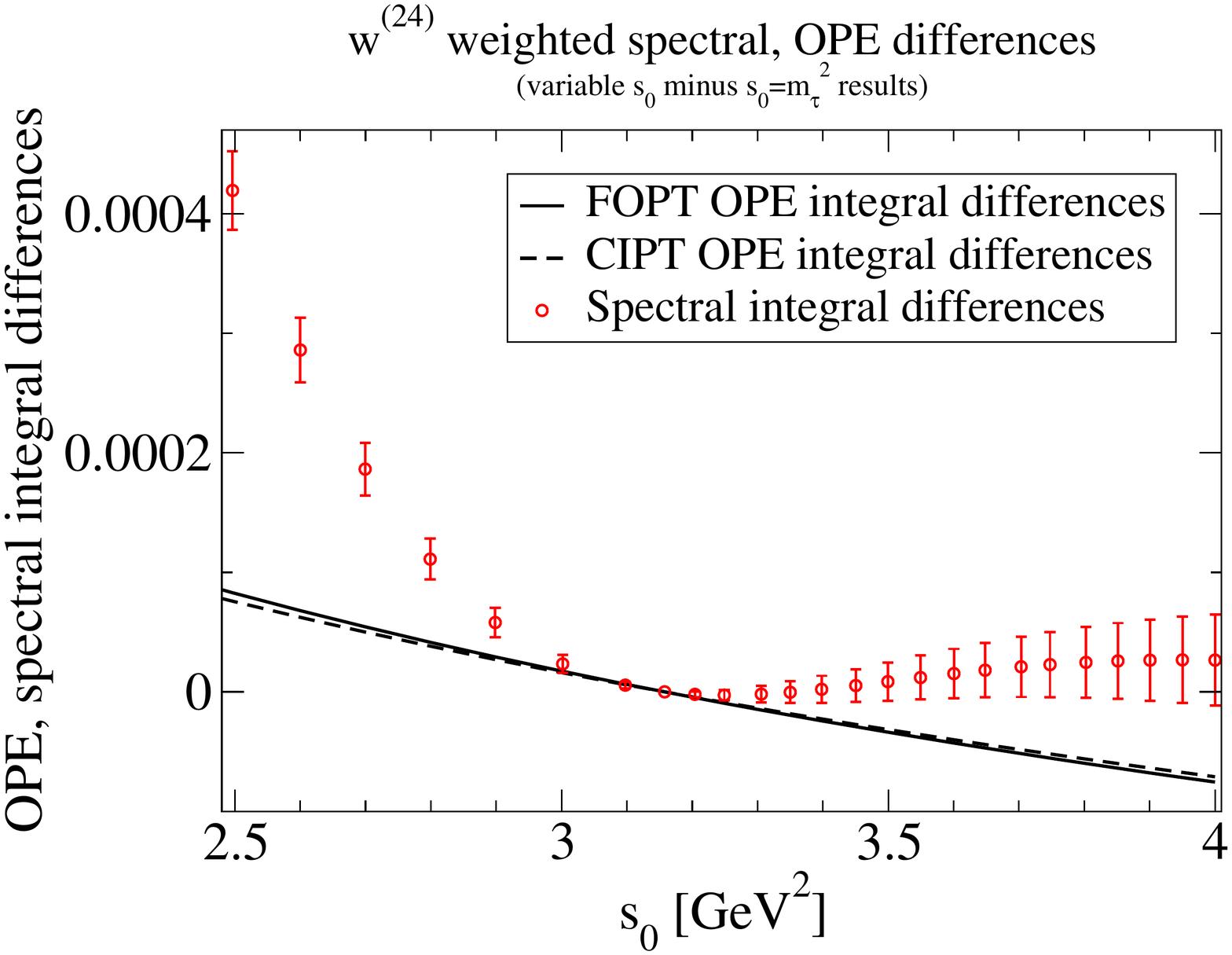}
\vspace{-2cm}
\begin{center}
\includegraphics*[width=10.5cm]{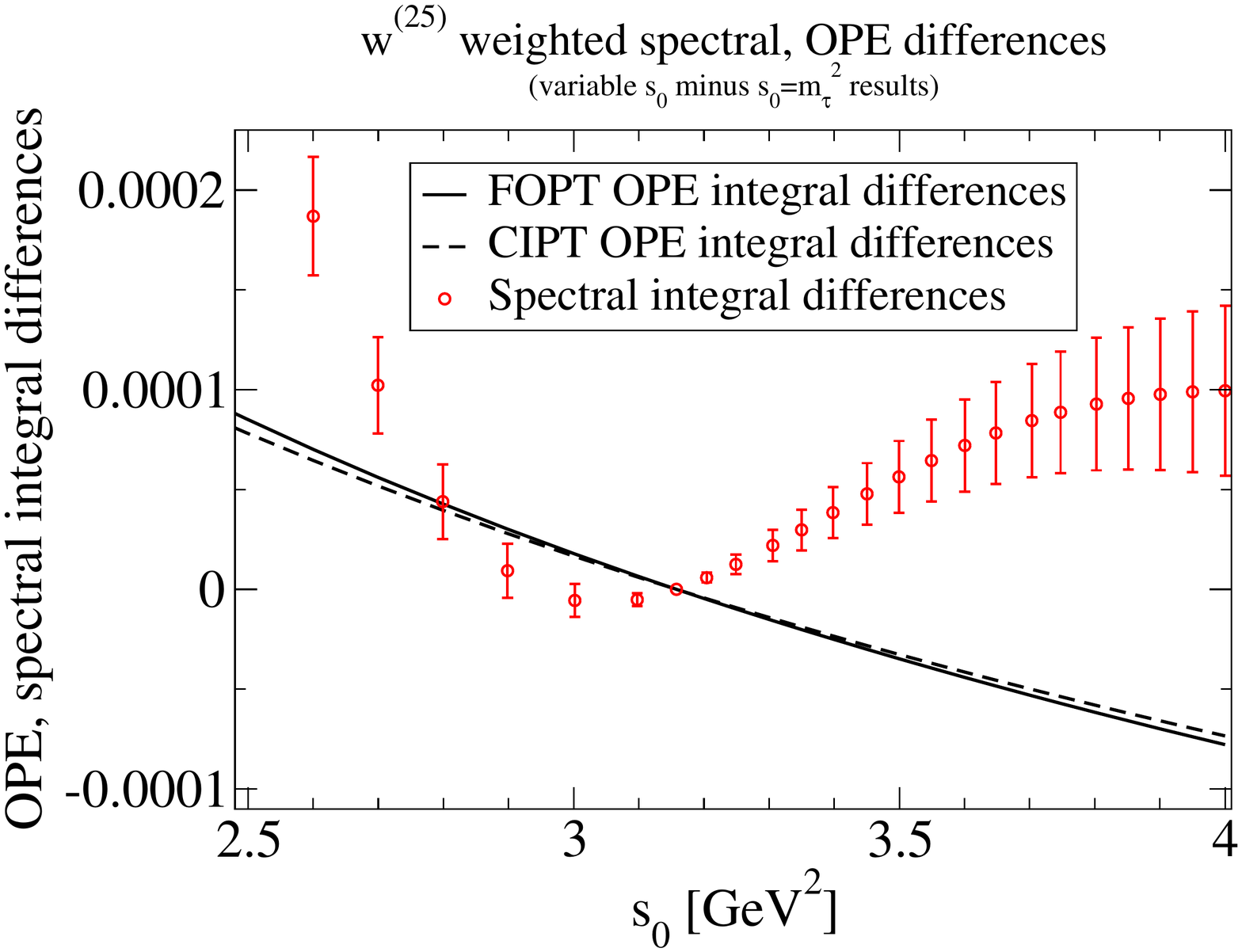}
\end{center}
\begin{quotation}
\caption{{\it EM FESR tests of
the optimal weight version of the truncated OPE strategy. Comparisons
of differences between general $s_0$ and $s_0=m_\tau^2$ versions
of the OPE and spectral integrals, with OPE results corresponding
to OPE parameter values obtained from the optimal-weight implementation
of the truncated OPE strategy using $s_0=m_\tau^2$ only in the fits.
Top left: $w^{(21)}$; top right: $w^{(22)}$; middle left: $w^{(23)}$;
middle right: $w^{(24)}$; bottom: $w^{(25)}$.
}}
\end{quotation}
\vspace*{-4ex}
\end{figure}

\section{$e^+e^-$-based tests of the truncated-OPE FESR strategy}
\label{sec:emtests}

In this section we focus our investigation of the assumptions
underlying the truncated OPE strategy on two of the sets of weights
employed in the nominal self-consistency studies of Ref.~\cite{Pich},
namely the conventional ``$(kl)$ spectral weight'' set,
\begin{equation}
w_{kl}(y)=y^l(1-y)^{k}\, w_\tau (y)\ ,
\label{wkldefn}\end{equation}
with $(kl)=(00)$, $(10)$, $(11)$, $(12)$ and $(13)$, and the set
of so-called ``optimal weights'',
\begin{equation}
w^{(2n)}(y)=1-(n+2)y^{n+1}+(n+1)y^{n+2}\ ,
\label{optwtdefn}\end{equation}
with $n=1,\cdots ,5$.

The $(00)$ spectral weight is doubly
pinched and the remainder of the {$(kl)$} spectral weights triply pinched,
while the optimal weights are all doubly pinched. Since both the $(kl)$ spectral
weight and optimal weight sets include weights up to degree $7$,
the corresponding sets of FESRs involve, in principle, OPE contributions,
unsuppressed by additional factors of $\alpha_s$, up to $D=16$. In
order to leave one more $s_0=m_\tau^2$ spectral integral than
OPE parameter in the corresponding multi-weight fits,
spectral-weight analyses fit $\alpha_s$, $C_4$, $C_6$ and $C_8$
and assume contributions proportional to $C_{10}$, $C_{12}$, $C_{14}$
and $C_{16}$ can be neglected. The absence of a term linear in $y$
in the weights $w^{(2n)}(y)$ means that contributions proportional
to $C_4$ are strongly suppressed. The five $s_0=m_\tau^2$
optimal-weight-set spectral integrals are then used to fit the four
OPE parameters, $\alpha_s$, $C_6$, $C_8$ and $C_{10}$, with contributions
proportional to $C_{12}$, $C_{14}$ and $C_{16}$ assumed negligible.

We consider spectral-weight and optimal-weight FESRs, in which the
$ud$ $V$ or $A$ spectral functions and HVPs appearing in Eq.~(\ref{basicfesr})
are replaced by the corresponding spectral function,
$\rho_{EM}(s)$, and HVP, $\Pi_{EM}(s)$, of the
three-flavor EM current. The spectral function, $\rho_{EM}(s)$,
is related to the well-known $R(s)$ ratio by
\begin{equation}
\rho_{EM}(s) ={\frac{1}{12\pi^2}}\, R(s)\ .
\label{rhoemcfR}\end{equation}
We employ the results and covariances for $R(s)$
provided by the authors of Ref.~\cite{knt2018}.
Full details of our {own} implementation of such EM FESRs may be found in
Ref.~\cite{bgkmnpt2018}.

We stress that (i) in the isospin limit,
CVC implies that {{the $I=1$ part of $\Pi_{EM}(s)$, $\Pi_{EM}^{I=1}$
is {equal to} ${\frac{1}{2}}\,\Pi_{ud;V}^{(0+1)}(s)$}},
where the $1/2$ is a trivial Clebsch-Gordon factor, and (ii) the
$I=0$ part of $\Pi_{EM}(s)$ is, up to a factor of $1/3$, the $SU(3)_F$
hypercharge partner of the $I=1$ component. Higher dimension $I=0$ EM
OPE condensate contributions to $\Pi_{EM}(s)$ should thus be
$\sim 1/3$ of the corresponding $I=1$ EM OPE condensate contributions,
up $SU(3)_F$ breaking effects. If $I=1$ condensates of a given dimension
{ yielded} contributions which are negligible, relative to perturbative
contributions, for $s_0\geq m_\tau^2$, this should be equally
true of the corresponding $I=0$ contributions. It follows that, if
the truncated-OPE-strategy assumptions {were} reliable at $s_0=m_\tau^2$
for $\tau$-decay-based FESRs, they should be similarly reliable
at $s_0=m_\tau^2$ for the corresponding EM FESRs, and they should
then be even more reliable for $s_0>m_\tau^2$, though this
expectation can only be tested in the EM case.

{Of course, the EM case allows us to consider only the $V$ channel, whereas
Ref.~\cite{Pich} considers $V+A$ to be the optimal choice for the
truncated OPE strategy. We note, however, that (i) there is not a
vast difference between the amplitude of the DV oscillations in the
$V$ and $V+A$ channels, relative to the
parton model, and (ii), that,
in particular for the optimal weights, the results for $\alpha_s$ obtained in
Ref.~\cite{Pich} on the basis of the truncated OPE strategy
are in excellent agreement between fits to the $V$ and $V+A$ channels,
while the corresponding agreement for the spectral
weights is also very good.}

We test the truncated OPE strategy by first performing
truncated-OPE-strategy fits to the $s_0=m_\tau^2$ versions
of either the five {$w_{kl}$}-weighted EM spectral integrals or
the five $w^{(2n)}$-weighted EM integrals, and then comparing
the weighted EM spectral integrals and OPE integrals obtained
using the resulting fitted OPE parameters at $s_0>m_\tau^2$.

Very strong correlations exist between weighted spectral integrals
for different $s_0$, as well as between weighted OPE integrals for
different $s_0$. In order to take these correlations into account
in assessing, visually, how successful the resulting $s_0>m_\tau^2$
OPE integrals are in predicting the actual values of the
corresponding EM spectral integrals, it is useful to plot
not the spectral and OPE integrals themselves, but rather
the difference between their values at general $s_0$ and
$s_0=m_\tau^2$. Both the OPE and spectral integral differences
are thus zero, by definition, at $s_0=m_\tau^2$. The errors
on the spectral integral differences are straightforwardly
obtainable from the covariance matrix of the $R(s)$ data provided
by the authors of Ref.~\cite{knt2018}.

The results of this test are shown in Fig.~\ref{optimalwttests},
for the optimal-weight set of Ref.~\cite{Pich}. The
OPE integral differences produced using the truncated-OPE-strategy fit
assumptions obviously provide an, in general, very poor representation
of the corresponding spectral integral differences in the region
above $s_0=m_\tau^2$. For the sake of brevity, the OPE-spectral
integral matches of the analogous spectral-weight test, which
are similarly bad above $s_0=m_\tau^2$, are not shown here.

From these results it is clear that the assumptions underlying
the truncated OPE strategy are, simply, {not valid, and thus that results
obtained from the truncated OPE strategy are unreliable}. Since
the weights involved in these
tests are doubly and/or triply pinched, and hence expected to
have suppressed integrated DV contributions, especially above
$s_0=m_\tau^2$, the poor OPE-spectral
integral matches imply a breakdown of the assumption that the OPE
 can be truncated as it would were the OPE a rapidly converging
 expansion up to at least $D=16$.

The consequences of this observation for $\tau$-based analyses
are (i) that the truncations in dimension of the OPE employed in
the truncated OPE strategy are completely unsafe and (ii) that,
in order to have fewer OPE parameters than spectral integrals
required to fit them, one must consider also spectral integrals
involving whatever set of weights one is employing at $s_0$
different from $m_\tau^2$, which, for analyses of {$\tau$-decay}
data, means $s_0<m_\tau^2$. Since quite sizeable DV oscillations
about perturbation theory are observed in the spectral functions
in this region, even when one considers the $ud$ $V+A$ sum, it
becomes important to use some representation of DV contributions
to estimate the impact of possible residual DV effects, even
in FESRs involving doubly and triply pinched weights.

\section{Duality Violations and Hyperasymptotics: The Regge Connection}
\label{sec:Regge}

Although one expects DVs to behave as in {Eq.~(\ref{parametrization})} for
large $s$ on general grounds, it would be {nice to derive} an
expression such as {Eq.~(\ref{parametrization})} from QCD. {Regrettably,}
this is still not possible from first principles but, recently,
in Ref.~\cite{Regge}, progress has been made under two plausible
assumptions: {(i) that} the radial spectrum of QCD shows a leading
Regge behavior in the vector channel for asymptotically large excitation
number $n$, {\it i.e.},
\begin{eqnarray}\label{regge}
  M^2(n) &=& \Lambda_{QCD}^2\, n+ b \log n + c +
\mathcal{O}\left( \frac{1}{n}, \frac{1}{\log n}\right)\ ,\\ \nn
  \frac{F(n)}{F_0}&=&1+\mathcal{O}\left( \frac{1}{n},
\frac{1}{\log n}\right)\ , \quad  \quad ( n\gg 1)\ ,\quad
\end{eqnarray}
where $M(n)$ is the spectrum of masses and $F(n)$ are the corresponding
decay constants appearing in the vector two-point function, in the
large-$N_c$ limit; and {(ii) that} the ratio of the width over the mass goes
to a constant also in the same asymptotic limit, {\it i.e.},
\begin{equation}\label{width}
  \frac{\Gamma}{M(n)} =\frac{a}{N_c}  \left(1+ \mathcal{O}
\left(\frac{1}{N_c},\frac{1}{n}\right)\right)\ , \quad  \quad ( n\gg 1)\ .
\end{equation}
The scale $\Lambda_{QCD}$ is related to the string tension
and is expected to be of order $1$ GeV (see also below). The scale $F_0$ sets the normalization of the two-point function.

Both assumptions are supported by  the solution of two-dimensional
QCD \cite{Blok}, the string picture of hadrons \cite{string} and
phenomenology \cite{Masjuan}. The picture that emerges is the following.

{Starting from} the dispersive {representation} obeyed by the
Adler function, it is convenient to express it as a Borel--Laplace transform

\bea
 \lbl{adler}
 \mathcal{A}(q^2)=-q^2 \frac{d\Pi(q^2)}{dq^2}
 &=& -\, q^2\int_0^{\infty}dt\ \rho(t) \int_0^{\infty}
d\sigma \ \sigma\ \mathrm{e}^{-\sigma\left(t-q^2\right)}\\
 &=& -\, q^2 \int_0^{\infty} d\sigma\ \mathrm{e}^{\sigma q^2}
\, \sigma \borel\ ,\nn
 \eea
where
 \be
 \lbl{rhohat}
  \borel=\int_0^{\infty}dt\  \rho(t)\  \mathrm{e}^{-\sigma t}
  \ee
is the Laplace transform of the spectral function. The OPE corresponds
to an expansion of $\borel$ around $\sigma=0$. We see that
$\borel$ is well-defined for Re~$\sigma>0$, since $\rho(t)$ ({{\it i.e.},} the
spectral function) must go to a constant as $t\to\infty$, for {finite $N_c$}. Any
singularities of $\borel$ thus have to reside in the
half-plane Re~$\sigma\le 0$. This representation of the
Adler function in terms of $\borel$ is valid for Re$(\sigma q^2)<0$,
and for $\sigma>0$ this means $q^2<0$. This is the key point: as one
rotates $\sigma$ in the complex plane from Re$\, \sigma>0$ to
Re$\, \sigma<0$, one is analytically continuing in the $q^2$
complex plane from $q^2<0$ to $q^2>0$. This  is what we want.

\begin{figure}
\label{fig:regge}
\begin{center}
\includegraphics*[width=14cm]{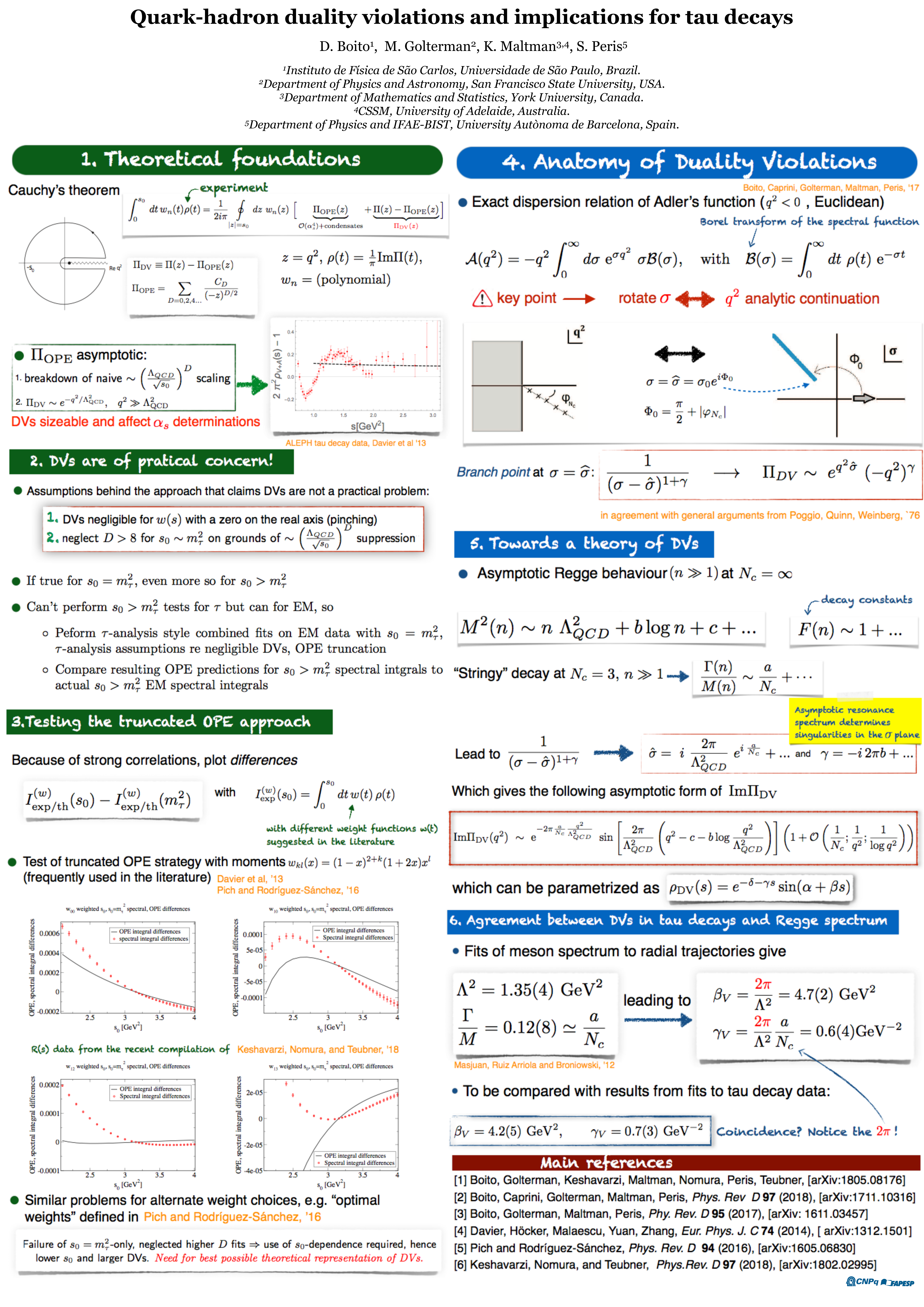}
\end{center}
\begin{quotation}
\caption{{\it  Schematic representation of the connection
between the singularities in the $q^2$ and $\sigma$ complex planes.
The thick gray arrow in the right panel depicts the initial path
taken in the $sigma$ integral in Eq. (\ref{adler}).}}
\end{quotation}
\end{figure}

If the spectrum $\rho(t)$ {were} to vanish for $t>t_0$, the
function $\borel$ would be analytic in the whole complex
plane, and the above rotation in $\sigma$ would produce an
OPE convergent for $|q^2|>t_0$. Of course, the spectrum goes
all the way to infinity, as Eq.~(\ref{regge}) clearly shows.
The existence of {an infinite number of} poles of {$\Pi(q^2)$} on the Minkowski
axis for $N_c=\infty$ produces singularities for $\borel$ on the
imaginary axis in the $\sigma$ plane which, for the spectrum
in {Eq.~(\ref{regge})}, are branch points \cite{Regge}. As $N_c$
evolves from infinity down to $3$, the location of these poles
recedes into the next Riemann sheet an angle given by
\begin{equation}\label{angle}
  \varphi_{N_c}=- \frac{\Gamma}{M(n)} =-\frac{a}{N_c}
\left(1+ \mathcal{O}\left(\frac{1}{N_c},\frac{1}{n}\right)\right) \ ,
\end{equation}
turning what were poles on the real $q^2>0$ axis into resonance peaks. Since $\sigma$ and
$q^2$ are locked together, through Eq.~(\ref{adler}),
to satisfy Re$(\sigma q^2)<0$,
this forces {all branch points in the $\sigma$ plane off the imaginary axis,
with the one closest to the origin moving} to a position given by
\begin{eqnarray}\label{anglesigma}
\sigma&=&\hat{\sigma}\approx \frac{2\pi}{\Lambda_{QCD}^2}\ e^{i \Phi_0}\ ,\nn \\
  \Phi_0&\approx &\frac{\pi}{2}+ |\varphi_{N_c}|\ .
\end{eqnarray}
The position of the branch point is signaled by a blue
arrow in Fig.~\ref{fig:regge}.

In this situation, as the path in the $\sigma$ plane is rotated in
the integral (\ref{adler}) from $\arg \sigma=0$ to $\arg \sigma=\pi$,
one sweeps through the blue line in Fig.~\ref{fig:regge}, picking up a
contribution given by
\begin{equation}\label{DVs}
  \mathrm{Im}\,\Pi_{DV}(q^2)\sim e^{\!\!\!-2\pi \frac{a}{N_c}
\frac{q^2}{\Lambda_{QCD}^2}}\ \!\! \sin\left[\frac{2\pi}
{\Lambda_{QCD}^2}\left( q^2-c-b \log\frac{q^2}{\Lambda_{QCD}^2}
\right)\right]\!\!\left(\!\!1+\!\!\mathcal{O}\left(\frac{1}{N_c},
\frac{1}{q^2},\frac{1}{\log q^2}\right)\!\!\!\right) \ .
\end{equation}
This expression can be parametrized as in {Eq.~(\ref{parametrization})},
up to a small logarithmic corrections (since, for large $q^2$,
$q^2\gg  b \log q^2$). In fact, QCD Regge phenomenology is consistent with
this term $b$ being absent.

Besides the branch point~(\ref{anglesigma}), in principle there may be other branch
points located in the same quadrant further away from the origin but, since the exponent in {Eq.~(\ref{DVs})} is governed by the radial distance of these points to the origin,
their contribution to Im\,$\Pi_{DV}$ will correspondingly contain a stronger exponential
suppression. In this way, the expansion at large $q^2$ of Im\,$\Pi_{DV}$
becomes a combined series in $1/q^2$ and exponentials $e^{-q^2}$, of
decreasing importance, as in the Theory of Hyperasymptotics \cite{math}.

Equation~(\ref{DVs}) connects the parameters from the radial Regge
trajectories~(\ref{regge}) to the parameters $\alpha_V,\, \beta_V,\, \gamma_V$
and $\delta_V$ of Eq.~(\ref{parametrization}), which were obtained  from fits involving the vector spectral function
in $\tau$ decay. On the other hand, fits {to meson} spectroscopy
give~\cite{Masjuan}\footnote{For example, in the case of the $\rho$,
one finds $\Gamma/M \simeq 0.19$.}
\begin{equation}\label{masjuanfit}
  \Lambda_{QCD}^2=1.35(4)\ \mathrm{GeV}^2\ , \quad \frac{\Gamma}{M}=0.12(8)\ ,
\end{equation}
which translate into
\begin{equation}\label{masjuanfit2}
  \beta_V=\frac{2\pi}{\Lambda_{QCD}^2}=4.7(2)\ \mathrm{GeV}^2\ ,
\quad \gamma_V=\frac{2\pi}{\Lambda_{QCD}^2}\frac{a}{N_c}=0.6(4)\
\mathrm{GeV}^{-2}\ .
\end{equation}
These numbers are to be compared to the results from the fit involving
{$\tau$} data \cite{tau}:
\begin{equation}\label{fittau}
  \beta_V=4.2(5)\ \mathrm{GeV}^{-2}\quad ,
\quad \gamma_V=0.7(3)\ \mathrm{GeV}^{-2}\ .
\end{equation}
The agreement is rather satisfactory. Notice in particular the importance
of having the factors of $2\pi$ in {Eq.~(\ref{masjuanfit2})}.

\section{Conclusion}
\label{sec:conclusions}

We have argued that the mass of the $\tau$ lepton is not high enough
to be able to dismiss the DV term {(\ref{pidvdefn}) in the FESR (\ref{basicfesr})} and that,
{because of that},
one has to use a parametrization of the DV term which is
physically sound, such as that given in {Eq.~(\ref{parametrization})}.
Attempts to work only at $s_0=m_\tau^2$, assuming integrated
DVs are negligible at this $s_0$ for doubly and triply pinched weights,
{run} into the problem that the number of OPE parameters to be fit
exceeds the number of spectral integrals available as input,
unless, as in the truncated OPE strategy, one neglects sufficiently
many {higher-$D$} OPE contributions present in the
analysis. We tested the reliability of the truncated OPE
strategy, which neglects such higher-$D$ contributions, using EM
FESRs employing recent $R(s)$ data as input, and found {that this strategy, and the assumptions
underlying it, fail} badly. This leads us to the conclusion that
one must take advantage of the $s_0$ dependence of $\tau$-based
spectral integrals to {have} enough input to fit all relevant OPE
parameters which, in turn, forces us to work at lower scales, where it
becomes {more} important to take DVs into account. We conclude that
an accurate extraction of $\alpha_s$ using {$\tau$-decay data
not} subject to uncontrolled systematic errors requires
a reasonable description of the DVs.

These conclusions {stand in sharp contrast} to the claims of Ref.~\cite{Pich}.
The authors of Ref.~\cite{Pich} claim that DVs are sufficiently
suppressed in the $ud$ $V+A$ two-point function to be able to neglect
them altogether when using doubly or triply pinched weights, and {when}
working at the highest available scale, $s_0=m_\tau^2$. The use of
such weights, with their higher degrees, however, forces the authors
of Ref.~\cite{Pich} to make strong assumptions about the behavior of the (asymptotic) OPE series, in particular, that contributions from {higher-$D$} condensates
present in the FESRs they employ can be neglected. These assumptions have been tested using the
analogous EM FESRs and found to fail badly.  This rules out  the truncated OPE strategy employed in Ref.~\cite{Pich} as a reliable method for use in {the hadronic $\tau$-decay
determination of $\alpha_s$.

\section*{Acknowledgements}
{MG and SP gratefully acknowledge}  interesting discussions with
R. Miravitllas on the mathematical theory of Hyperasymptotics and
its connection to Duality Violations. DB, KM and SP would like to thank
the Department of Physics and Astronomy at San Francisco State
University for hospitality.  The work of D.B. is supported by the S\~ao Paulo
Research Foundation
(Fapesp) grant No. 2015/20689-9 and by the Brazilian National Council
for Scientific and Technological Development (CNPq), grant No.
305431/2015-3.
This material is based upon work supported by the U.S. Department
of Energy, Office of Science, Office of High Energy Physics, under
Award Number DE-SC0013682 (MG). KM is supported by a grant
from the Natural Sciences and Engineering Research Council of Canada,
and SP by
CICYTFEDER-FPA2014-55613-P, 2014-SGR-1450 and the
CERCA Program/Generalitat de Catalunya.

\nolinenumbers
}

\end{document}